\documentclass[aps,pra,twocolumn,superscriptaddress,nopacs,nofootinbib]{revtex4-1}

\usepackage{amsmath,amssymb,latexsym,amsthm}
\usepackage{graphicx}
\graphicspath{{../fig/}}
\usepackage{hyperref}
\usepackage{txfonts}
\usepackage{mathrsfs}
\usepackage{color}
\DeclareSymbolFont{symbols}{OMS}{cmsy}{m}{n}
\DeclareSymbolFont{largesymbols}{OMX}{cmex}{m}{n}
\renewcommand{\bm}[1]{\boldsymbol #1}

\begin{document}

\title{
Bound on the exponential growth rate of out-of-time-ordered correlators
}

\author{Naoto Tsuji}
\affiliation{RIKEN Center for Emergent Matter Science (CEMS), Wako 351-0198, Japan}
\author{Tomohiro Shitara}
\affiliation{Department of Physics, University of Tokyo, Hongo, Tokyo 113-0033, Japan}
\author{Masahito Ueda}
\affiliation{Department of Physics, University of Tokyo, Hongo, Tokyo 113-0033, Japan}
\affiliation{RIKEN Center for Emergent Matter Science (CEMS), Wako 351-0198, Japan}

\begin{abstract}
It has been conjectured by Maldacena, Shenker, and Stanford [J. High Energy Phys.~08 (2016) 106]
that the exponential growth rate of the out-of-time-ordered correlator (OTOC)
$F(t)$ has a universal upper bound $2\pi k_B T/\hbar$.
Here we introduce a one-parameter family of out-of-time-ordered correlators $F_\gamma(t)$
($0\leq\gamma\leq 1$), which has as good properties as $F(t)$ as a regularization of 
the out-of-time-ordered part of the squared commutator
$\langle [\hat A(t),\hat B(0)]^2\rangle$ that
diagnoses quantum many-body chaos,
and coincides with $F(t)$ at $\gamma=1/2$.
We rigorously prove that
if $F_\gamma(t)$ shows a transient exponential growth for all $\gamma$ in $0\leq\gamma\leq 1$,
that is, if the OTOC shows an exponential growth regardless of the choice of the regularization,
then the growth rate $\lambda$ does not depend on the regularization parameter $\gamma$,
and satisfies the inequality $\lambda\leq 2\pi k_B T/\hbar$.
\end{abstract}


\date{\today}


\maketitle

\section{Introduction}

Chaos in classical systems
is characterized by the Lyapunov exponent, which represents the maximal exponential growth rate of the distance
between different classical orbits that initially lie in the immediate vicinity of each other in the phase space.
The phenomenon that a tiny change in the initial condition blows up exponentially in time is known as a butterfly effect.
A quantum analog of the Lyapunov exponent, that has attracted much interest recently, is given by
an out-of-time-ordered correlator (OTOC) \cite{LarkinOvchinnikov1969}, a four-point correlation function 
such as $\langle \hat A(t)\hat B(0)\hat A(t)\hat B(0)\rangle$ that does not obey the usual time-ordering rule.
The motivation to consider such a correlator comes from the squared commutator
$-\langle [\hat A(t),\hat B(0)]^2\rangle$, which is a second moment of the variation of the operator $\hat A$ at time $t$
against a perturbation of a force $\hat B$ at time $0$,
and represents the sensitivity of the time-evolving observable to the initial perturbation.
If the OTOC grows exponentially in time,
one would expect that the growth rate of the OTOC plays a role similar to that of the Lyapunov exponent in quantum many-body systems \cite{Kitaev2015,MaldacenaShenkerStanford2016}.

Recently, a remarkable conjecture has been made by Maldacena, Shenker, and Stanford (MSS)
\cite{MaldacenaShenkerStanford2016}, stating that in thermal equilibrium
there exists a universal upper bound on the exponential growth rate $\lambda$
of the OTOCs,
\begin{align}
\lambda
\le
\frac{2\pi k_BT}{\hbar},
\label{MSS bound}
\end{align}
where $k_B$ is the Boltzmann constant, $T$ is the temperature of the system,
and $\hbar$ is the Planck constant. Precisely speaking, they introduce an OTOC of the form,
\begin{align}
F(t)
&\equiv
{\rm Tr}[\hat\rho^{\frac{1}{4}}\hat A(t)\hat\rho^{\frac{1}{4}}\hat B(0)
\hat\rho^{\frac{1}{4}}\hat A(t)\hat\rho^{\frac{1}{4}}\hat B(0)],
\label{F(t)}
\end{align}
where $\hat\rho=e^{-\beta\hat H}/Z$ is the thermal density-matrix operator, $\beta=1/k_BT$ is the inverse temperature,
$\hat H$ is the Hamiltonian,
$Z={\rm Tr}(e^{-\beta\hat H})$ is the partition function, and $\hat A$ and $\hat B$ are arbitrary hermitian operators. 
They focus on a situation where
there is a clear separation between the time scale (dissipation time) at which a usual time-ordered correlator decays to a constant
and the time scale (scrambling time) at which an OTOC grows exponentially. Let us suppose that
the OTOC (\ref{F(t)}) shows an exponential growth $F(t)=c_0-\epsilon c_1 e^{\lambda t}+O(\epsilon^2)$
($c_1, \lambda>0$)
with $t$ much after the dissipation time.
Here $\epsilon$ is a certain small positive expansion parameter such as $\hbar^2$ in the semiclassical approximation
or $1/N^2$ in large-$N$ theories. Then the MSS conjecture states that $\lambda$ always satisfies the inequality
(\ref{MSS bound}) regardless of the choice of $\hat A$ and $\hat B$ and the details of $\hat H$.
In this sense, the bound is completely universal, and is thought to be a fundamental property of quantum systems.
It may be viewed as a refinement of the fast scrambling conjecture \cite{SekinoSusskind2008}.
Several examples are known to saturate the bound (\ref{MSS bound}), including black holes in Einstein gravity
\cite{ShenkerStanford2014a,ShenkerStanford2014b,ShenkerStanford2015,
MaldacenaShenkerStanford2016}
and the Sachdev-Ye-Kitaev model \cite{SachdevYe1993,Kitaev2015,PolchinskiRosenhaus2016,MaldacenaStanford2016}.
Various analytical as well as numerical calculations have been performed 
for the growth rate of OTOCs in many different systems \cite{RobertsStanford2015,Stanford2016,RobertsSwingle2016,Hashimoto2016,Yao2016,Rozenbaum2016,Banerjee2017, 
PatelSachdev2016,Kurchan2016,Bohrdt2016,ChowdhurySwingle2017,Patel2017}.
No clear counterexample that violates the bound (\ref{MSS bound}) has been presented so far.

The motivation to consider $F(t)$ in Eq.~(\ref{F(t)}) is that in quantum field theory
$\langle [\hat A(t),\hat B(0)]^2\rangle={\rm Tr}(\hat\rho [\hat A(t),\hat B(0)]^2)$ 
is not necessarily well-defined since two operators can approach in time arbitrarily close to each other.
A convenient prescription is to regularize it into 
${\rm Tr}(\hat\rho^{\frac{1}{2}}[\hat A(t),\hat B(0)]\hat\rho^{\frac{1}{2}}[\hat A(t),\hat B(0)])$ \cite{MaldacenaShenkerStanford2016}, which is called the bipartite OTOC \cite{TsujiShitaraUeda2016}.
In fact, the two are related to each other up to the difference of the Wigner-Yanase
skew information \cite{WignerYanase1963,TsujiShitaraUeda2016}, 
which is an information-theoretic measure of quantum fluctuations.
In the semiclassical regime of interest, the difference of the skew information is expected to be suppressed.
The out-of-time-ordered ($ABAB$ and $BABA$) part of the regularized OTOC is defined by
\begin{align}
F_0(t)
&\equiv
\frac{1}{2}
{\rm Tr}[\hat\rho^{\frac{1}{2}}\hat A(t)\hat B(0)
\hat\rho^{\frac{1}{2}}\hat A(t)\hat B(0)]
\notag
\\
&\quad
+\frac{1}{2}
{\rm Tr}[\hat\rho^{\frac{1}{2}}\hat B(0)\hat A(t)
\hat\rho^{\frac{1}{2}}\hat B(0)\hat A(t)].
\label{F_0(t)}
\end{align}
$F(t)$ in Eq.~(\ref{F(t)}) may be viewed as a variant of the regularization of the out-of-time-ordered part of the squared commutator.

The growth of the commutator is bounded by the Lieb-Robinson bound
\cite{LiebRobinson1972,Nachtergaele2006,Hastings2006,Hastings2010}, 
which gives a fundamental limit on the spread of information:
$\Vert[\hat A_{\bm x}(t),\hat B_{\bm y}(0)]\Vert\le c\Vert\hat A\Vert\, \Vert\hat B\Vert e^{-(|\bm x-\bm y|-vt)/\xi}$.
Here $\hat A$ and $\hat B$ are local operators inserted at positions $\bm x$ and $\bm y$, respectively,
$\Vert\cdot\Vert$ represents the operator norm, and $c$, $v$, and $\xi$ are some constants.
In contrast to the growth rate $v/\xi$ in the Lieb-Robinson bound,
the conjectured bound for $\lambda$ (\ref{MSS bound}) 
depends on the state of the quantum system, and the state dependence of the bound appears only through
the thermodynamic temperature
(the relation between the Lieb-Robinson bound and the quantum butterfly effect has been discussed in 
Refs.~\cite{RobertsStanfordSusskind2015,RobertsSwingle2016,Huang2016}).
Thus the bound (\ref{MSS bound}) constitutes a novel fundamental limit on the growth of information in general quantum systems.

A compelling argument has been given in Ref.~\cite{MaldacenaShenkerStanford2016}
to establish the conjecture (\ref{MSS bound}). The original derivation 
uses analytic properties of $F(z)$ (analytic continuation of $F(t)$ to complex time $z$) 
and a factorization of certain time-ordered correlation functions,
the latter of which has not been proved but used as a physical input \cite{MaldacenaShenkerStanford2016}.
The purpose of the present work is 
to rigorously prove (without assuming the factorization) 
that the inequality (\ref{MSS bound}) holds if the OTOC shows a transient exponential growth
in a certain time range
{\it irrespective of the way to regularize the squared commutator}.
In fact, there are not only $F(t)$ in Eq.~(\ref{F(t)}) and $F_0(t)$ in Eq.~(\ref{F_0(t)})
but also many other ways to regularize $\langle [\hat A(t),\hat B(0)]^2\rangle$.
Here we introduce a one-parameter family of OTOCs $F_\gamma(t)$ ($0\le\gamma\le 1$) 
[see Eq.~(\ref{one-parameter family})]
that interpolates between $F_0(t)=F_{\gamma=0}(t)=F_{\gamma=1}(t)$ and $F(t)=F_{\gamma=\frac{1}{2}}(t)$.
We show that $F_\gamma(t)$ has as good properties as $F(t)$ (\ref{F(t)}) and $F_0(t)$ (\ref{F_0(t)})
as a regularization of the out-of-time-ordered part of $\langle [\hat A(t),\hat B(0)]^2\rangle$.
If the exponential growth of the OTOC is physically meaningful (or universal), 
it should not depend on the choice of the regularization.
Hence it is reasonable to require that all the members in the one-parameter family of the OTOCs $F_\gamma(t)$
($0\le\gamma\le 1$) grow exponentially in time.
Under this requirement, we rigorously prove the existence of the bound (\ref{MSS bound})
on the exponential growth rate of the OTOCs.

The rest of the paper is organized as follows.
In Sec.~\ref{OTOC family}, we introduce a one-parameter family of OTOCs that makes
as much sense as $F(t)$ (\ref{F(t)}) and $F_0(t)$ (\ref{F_0(t)}) as a regularization of the squared commutator.
In Sec.~\ref{theorem}, we describe the statement of the main theorem in this paper that claims the existence
of the bound on the exponential growth rate of OTOCs, and prove it.
In Sec.~\ref{discussion}, we discuss various issues related to the theorem, including
a generalization of the theorem to higher-order OTOCs.

\section{One-parameter family of OTOCs}
\label{OTOC family}

We introduce a one-parameter family of OTOCs
\begin{align}
F_\gamma(t)
&\equiv
\frac{1}{2}
{\rm Tr}[\hat\rho^{\frac{1-\gamma}{2}}\hat A(t)\hat\rho^{\frac{\gamma}{2}}\hat B(0)
\hat\rho^{\frac{1-\gamma}{2}}\hat A(t)\hat\rho^{\frac{\gamma}{2}}\hat B(0)]
\notag
\\
&\quad
+\frac{1}{2}
{\rm Tr}[\hat\rho^{\frac{1-\gamma}{2}}\hat B(0)\hat\rho^{\frac{\gamma}{2}}\hat A(t)
\hat\rho^{\frac{1-\gamma}{2}}\hat B(0)\hat\rho^{\frac{\gamma}{2}}\hat A(t)]
\label{one-parameter family}
\end{align}
for $0\le\gamma\le 1$. We note that
$F_\gamma(t)$ is symmetric around $\gamma=\frac{1}{2}$ (i.e., $F_\gamma(t)=F_{1-\gamma}(t)$),
and agrees with $F_0(t)$ (\ref{F_0(t)}) at $\gamma=0,1$ and $F(t)$ (\ref{F(t)})
at $\gamma=\frac{1}{2}$.
This form of the OTOC has appeared in the study of the out-of-time-order fluctuation-dissipation theorem
\cite{TsujiShitaraUeda2016}. If one defines 
\begin{align}
C_{[A,B]_{\alpha_1}[A,B]_{\alpha_2}}^\gamma(t,0)
&\equiv
{\rm Tr}[\hat\rho^{\frac{1-\gamma}{2}}(\hat A(t)\hat\rho^{\frac{\gamma}{2}}\hat B(0)+\alpha_1\hat B(0)\hat\rho^{\frac{\gamma}{2}}\hat A(t))
\notag
\\
&\quad\times
\hat\rho^{\frac{1-\gamma}{2}}(\hat A(t)\hat\rho^{\frac{\gamma}{2}}\hat B(0)+\alpha_2\hat B(0)\hat\rho^{\frac{\gamma}{2}}\hat A(t))],
\label{C^gamma}
\end{align}
where $\alpha_1,\alpha_2=\pm$ and $[,]_{-(+)}=[,]$ ($\{,\}$) is the (anti)commutator, then
$4F_\gamma(t)=C_{\{A,B\}^2}^\gamma(t,0)+C_{[A,B]^2}^\gamma(t,0)$
coincides with the left-hand side of the out-of-time-order fluctuation-dissipation theorem
\cite{TsujiShitaraUeda2016},
\begin{align}
&
C_{\{A,B\}^2}^\gamma(\omega)+C_{[A,B]^2}^\gamma(\omega)
\notag
\\
&\qquad=
2\coth\left((1-2\gamma)\frac{\beta\hbar\omega}{4}\right)C_{\{A,B\}[A,B]}^\gamma(\omega).
\end{align}
Here $C_{[A,B]_{\alpha_1}[A,B]_{\alpha_2}}^\gamma(\omega)
\equiv\int_{-\infty}^{\infty} dt\, e^{i\omega t}C_{[A,B]_{\alpha_1}[A,B]_{\alpha_2}}^\gamma(t,0)$
is the Fourier transform of Eq.~(\ref{C^gamma}).
In other words, $F_\gamma(t)$ corresponds to the ``fluctuation'' part of 
the fluctuation-dissipation relation.

\begin{figure}[t]
\includegraphics[width=7.5cm]{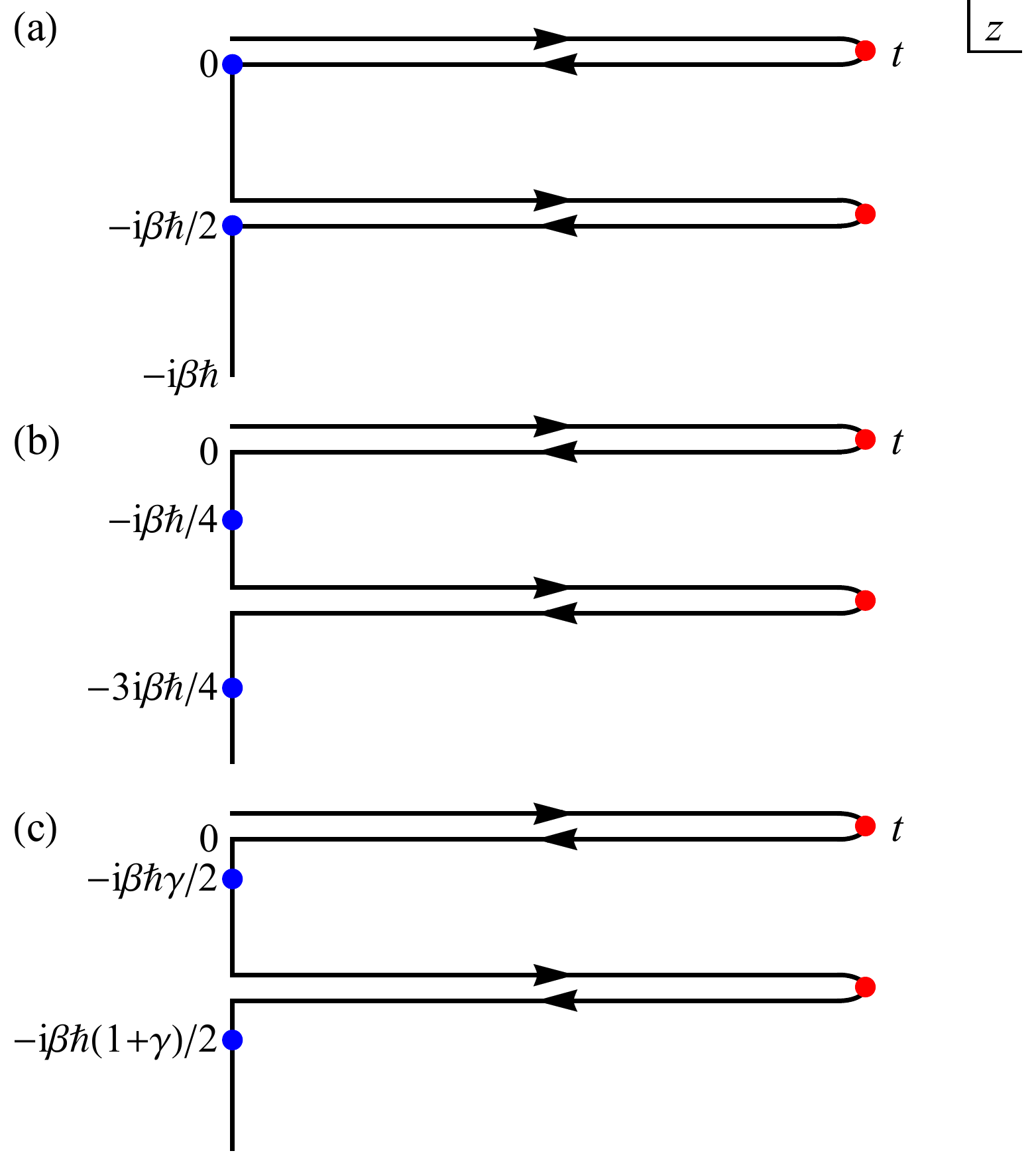}
\caption{Positions of the operators inserted along the contour $\mathcal C$ 
in the complex time domain for (a) the bipartite OTOC $F_0(t)$,
(b) the symmetric OTOC $F(t)=F_{\frac{1}{2}}(t)$, and (c) the generalized OTOC $F_\gamma(t)$ ($0\le\gamma\le 1$).
Red (blue) dots represent the positions of the operators $\hat A(z)$ ($\hat B(z)$).}
\label{contour}
\end{figure}

Each term in the OTOCs can be represented as a contour-ordered function 
\begin{align}
\frac{1}{Z}
{\rm Tr}[\mathcal T_{\mathcal C} e^{-\frac{i}{\hbar}\int_{\mathcal C} dz \hat H(z)}\hat A(z_1)\hat B(z_2)\hat A(z_3)\hat B(z_4)]
\quad
(z_i\in\mathbb C),
\end{align}
where the contour $\mathcal C$ has double-folded branches \cite{MaldacenaShenkerStanford2016,Aleiner2016,
TsujiWernerUeda2016,Haehl2016a}
in the complex time domain as depicted in Fig.~\ref{contour},
and $\mathcal T_{\mathcal C}$ is the time ordering operator along $\mathcal C$.
The positions of the operators inserted along the contour $\mathcal C$ are shown for $F_0(t)$, $F(t)=F_{\frac{1}{2}}(t)$,
and $F_\gamma(t)$ in Figs.~\ref{contour}(a), (b), and (c), respectively.

The one-parameter family of the OTOC $F_\gamma(t)$ ($0\le\gamma\le 1$) (\ref{one-parameter family})
has as good properties as $F(t)$ as a regularization of the out-of-time-ordered part of $\langle [\hat A(t),\hat B(0)]^2\rangle$. 
First, $F_\gamma(t)$ is real if $\hat A$ and $\hat B$ are hermitian.
Hence it makes sense to discuss the sign of the variation of $F_\gamma(t)$, which plays an important role below.
Second, $F_\gamma(t)$ smoothly interpolates between $F_0(t)$ (\ref{F_0(t)}) and $F(t)=F_{\frac{1}{2}}(t)$ (\ref{F(t)}),
corresponding to the continuous shift of the positions of the operators inserted on the imaginary-time axis
from Fig.~\ref{contour}(a) to (b) through (c).
Third, $F_\gamma(t)$ is the out-of-time-ordered ($ABAB$ and $BABA$) part of
$\frac{1}{2}C_{[A,B]^2}^\gamma(t,0)$,
which is a kind of generalization of the squared commutator.
If one defines a generalized commutator as
\begin{align}
[\hat A,\hat B]_\gamma\equiv\hat A\hat\rho^{\frac{\gamma}{2}}\hat B-\hat B\hat\rho^{\frac{\gamma}{2}}\hat A
\quad
(0\le\gamma\le 1),
\end{align}
then the bracket $[\cdot,\cdot]_\gamma$ satisfies the bilinearity,
$[a\hat A+b\hat B,\hat C]_\gamma=a[\hat A,\hat C]_\gamma+b[\hat B,\hat C]_\gamma$
and $[\hat C,a\hat A+b\hat B]_\gamma=a[\hat C,\hat A]_\gamma+b[\hat C,\hat B]_\gamma$ ($a,b\in\mathbb C$),
the alternativity $[\hat A,\hat A]_\gamma=0$, and the Jacobi identity, 
$[\hat A,[\hat B,\hat C]_\gamma]_\gamma+[\hat B,[\hat C,\hat A]_\gamma]_\gamma+[\hat C,[\hat A,\hat B]_\gamma]_\gamma=0$.
Hence the bracket $[\cdot,\cdot]_\gamma$ satisfies the axiom of the commutator (or the Lie algebra).
If $\hat A$ and $\hat B$ are hermitian, then the generalized commutator $[\hat A,\hat B]_\gamma$ is skew-hermitian,
i.e., $([\hat A,\hat B]_\gamma)^\dagger=-[\hat A,\hat B]_\gamma$.
$C_{[A,B]^2}^\gamma(t,0)$ can be expressed as
$C_{[A,B]^2}^\gamma(t,0)
=
{\rm Tr}(\hat\rho^{\frac{1-\gamma}{2}}[\hat A(t),\hat B(0)]_\gamma
\hat\rho^{\frac{1-\gamma}{2}}[\hat A(t),\hat B(0)]_\gamma)$,
which contains two generalized commutators.
Since $C_{[A,B]^2}^\gamma(t,0)$ can be viewed as the trace of the square of the skew-hermitian operator,
it is negative semidefinite, $C_{[A,B]^2}^\gamma(t,0)\le 0$, as is the case
for the squared commutator $\langle [\hat A(t),\hat B(0)]^2\rangle$.
Therefore, if $C_{[A,B]^2}^\gamma(t,0)$ grows exponentially in such a manner
that the initial-perturbation sensitivity increases,
it should grow to the negative direction. Since the exponential growth of our interest
arises from the out-of-time-ordered ($ABAB$ and $BABA$) part \cite{MaldacenaShenkerStanford2016},
$F_\gamma(t)$ in Eq.~(\ref{one-parameter family}) should also grow to the negative direction. This is 
why we require
that $F_\gamma(t)=c_0(\gamma)-\epsilon c_1(\gamma) e^{\lambda(\gamma) t}+O(\epsilon^2)$ 
with $c_1(\gamma)\ge 0$ for $0\le\gamma\le 1$.

\section{Bound on the exponential growth rate of OTOCs}
\label{theorem}

Now we describe the statement of the main theorem that gives the rigorous bound on the exponential growth rate 
for the OTOCs, and prove it in two ways: One is to use a differential equation, and the other is
to use analytic continuation.

{\it Theorem. ---}
If the one-parameter family of the OTOC $F_\gamma(t)$ ($0\le\gamma\le 1$) (\ref{one-parameter family})
for hermitian operators $\hat A$ and $\hat B$
has a uniform asymptotic expansion of
\begin{align}
F_\gamma(t)
&=
c_0(\gamma)-\epsilon c_1(\gamma)e^{\lambda(\gamma) t}+O(\epsilon^2)
\label{asymptotic expansion}
\end{align}
in the region $D=\{(t, \gamma)\, | \, 0<t_1\leq t \leq t_2\, (t_1\neq t_2), 0\leq\gamma\leq 1\}$ with
$c_1(\gamma)\geq 0$ and $\lambda(\gamma)>0$
($0\leq \gamma\leq 1$), and if $c_1(\gamma)$ is nonzero at least at one $\gamma$ in $0\leq \gamma \leq 1$, then
the following properties hold:

(i) The exponent $\lambda(\gamma)$ is independent of $\gamma$ (hence we write $\lambda(\gamma)=\lambda$).

(ii) The coefficient $c_1(\gamma)$ is fully determined as
\begin{align}
c_1(\gamma)
&=
c_1 \cos\left((1-2\gamma)\frac{\beta\hbar\lambda}{4}\right)
\quad
(0\leq \gamma \leq 1)
\end{align}
with $c_1>0$.

(iii) The exponent $\lambda$ satisfies the inequality
\begin{align}
\lambda
\leq
\frac{2\pi}{\beta\hbar}
=
\frac{2\pi k_B T}{\hbar}.
\label{lambda bound}
\end{align}

Some technical remarks are in order. 
In the theorem,
we assume not only that $F_\gamma(t)$ has an asymptotic expansion of 
the form of Eq.~(\ref{asymptotic expansion}),
but also that the asymptotic expansion is {\it uniform}, that is, the speed of the convergence
of the expansion does not depend on $t$ and $\gamma$ in $D$. More precisely, 
$F_\gamma(t)$ converges to $c_0(\gamma)$ uniformly in $D$ in the limit of $\epsilon\to 0$,
and $(F_\gamma(t)-c_0(\gamma))/\epsilon$ converges to $-c_1(\gamma)e^{\lambda(\gamma)t}$
uniformly in $D$ in the limit of $\epsilon\to 0$. 
The assumption of uniform convergence is physically natural, since there is no {\it a priori} reason
that the convergence slows down at certain $t$ and $\gamma$ in the finite region $D$.
In the theorem, we exclude the trivial case in which $c_1(\gamma)$ vanishes for all $\gamma$
in $0\leq\gamma\leq 1$, since in this case $F_\gamma(t)$ does not show an exponential growth at all,
which is not of our interest here.

{\it Proof. ---} Let us write
\begin{align}
F_\gamma(t)
&=
\frac{1}{2}
{\rm Tr}\big[\hat\rho^{\frac{1}{4}}\hat A\big(t-i(\gamma-\tfrac{1}{2})\tfrac{\beta\hbar}{2}\big)\hat\rho^{\frac{1}{4}}\hat B(0)
\notag
\\
&\quad\times
\hat\rho^{\frac{1}{4}}\hat A\big(t-i(\gamma-\tfrac{1}{2})\tfrac{\beta\hbar}{2}\big)\hat\rho^{\frac{1}{4}}\hat B(0)\big]
\notag
\\
&\quad
+\frac{1}{2}
{\rm Tr}\big[\hat\rho^{\frac{1}{4}}\hat A\big(t+i(\gamma-\tfrac{1}{2})\tfrac{\beta\hbar}{2}\big)\hat\rho^{\frac{1}{4}}\hat B(0)
\notag
\\
&\quad\times
\hat\rho^{\frac{1}{4}}\hat A\big(t+i(\gamma-\tfrac{1}{2})\tfrac{\beta\hbar}{2}\big)\hat\rho^{\frac{1}{4}}\hat B(0)\big]
\notag
\\
&=
\frac{1}{2}F\big(t-i(\gamma-\tfrac{1}{2})\tfrac{\beta\hbar}{2}\big)
+\frac{1}{2}F\big(t+i(\gamma-\tfrac{1}{2})\tfrac{\beta\hbar}{2}\big).
\label{F_gamma(t) 1}
\end{align}
If we denote $z=t+i(\gamma-\frac{1}{2})\frac{\beta\hbar}{2}$, then $F_\gamma(t)$ can be expressed as
\begin{align}
F_\gamma(t)
&=
\frac{1}{2}F(z)+\frac{1}{2}F(\bar z).
\label{F_gamma(t) 2}
\end{align}
Since $F(\bar z)$ is the complex conjugate of $F(z)$ (i.e., $\overline{F(\bar z)}=F(z)$),
$F_\gamma(t)$ is the real part of the complex function $F(z)$.
Let us define $c_0 \equiv c_0(\frac{1}{2})$, $c_1 \equiv c_1(\frac{1}{2})\ge 0$, and $\lambda \equiv \lambda(\frac{1}{2})>0$.
At $\gamma=\frac{1}{2}$, we have
\begin{align}
F_{\frac{1}{2}}(t)
&=
F(t)
=
c_0-\epsilon c_1 e^{\lambda t}+O(\epsilon^2)
\quad
(t_1\le t \le t_2).
\label{F_1/2(t)}
\end{align}

It has been shown in Ref.~\cite{MaldacenaShenkerStanford2016} 
that $F(z)$ is analytic in the half strip region ${\rm Re}\,z>0$ and $-\frac{\beta\hbar}{4}\le{\rm Im}\,z\le\frac{\beta\hbar}{4}$
\cite{analyticity},
and especially in the region of 
$\Omega\equiv\{z\in\mathbb C \,|\, 0<t_1\le {\rm Re}\,z \le t_2, -\frac{\beta\hbar}{4}\le{\rm Im}\,z\le\frac{\beta\hbar}{4}\}$.
Hence $F(t)$ is infinitely differentiable,
and can be Taylor expanded around $t$ with the convergence radius of $\frac{\beta\hbar}{4}$.
This allows us to rewrite Eq.~(\ref{F_gamma(t) 1}) into a form of the differential equation,
\begin{align}
F_\gamma(t)
&=
\frac{1}{2}e^{-\frac{\beta\hbar}{2}(\gamma-\frac{1}{2})i\partial_t}F(t)
+\frac{1}{2}e^{\frac{\beta\hbar}{2}(\gamma-\frac{1}{2})i\partial_t}F(t)
\notag
\\
&=
\cos\left((1-2\gamma)\frac{\beta\hbar}{4}\partial_t\right)F(t).
\label{diff eq}
\end{align}
If $F(t)$ has the uniform asymptotic expansion (\ref{F_1/2(t)}) in $t_1\le t\le t_2$,
arbitrary-order derivatives of $F(t)$ also have uniform asymptotic expansions in $t_1\le t \le t_2$
since $F(z)$ is holomorphic in $\Omega$.
Therefore, we can exchange the derivative $\partial_t$ and the limit $\epsilon\to 0$ in Eq.~(\ref{diff eq}),
obtaining
\begin{align}
F_\gamma(t)
&=
c_0-\epsilon c_1 \cos\left((1-2\gamma)\frac{\beta\hbar\lambda}{4}\right)e^{\lambda t}+O(\epsilon^2).
\end{align}
This completely determines
the $\gamma$ dependences of $c_0(\gamma)$, $c_1(\gamma)$, and $\lambda(\gamma)$.
Especially, $\lambda(\gamma)=\lambda$ (independent of $\gamma$) and 
$c_1(\gamma)=c_1\cos\big((1-2\gamma)\frac{\beta\hbar\lambda}{4}\big)$.
If $c_1=0$, $c_1(\gamma)$ should vanish for all $\gamma$ in $0\le \gamma\le 1$,
which contradicts the assumption of the theorem.
Hence $c_1>0$, and the statements (i) and (ii) of the theorem follow.

\begin{figure}[t]
\includegraphics[width=7.5cm]{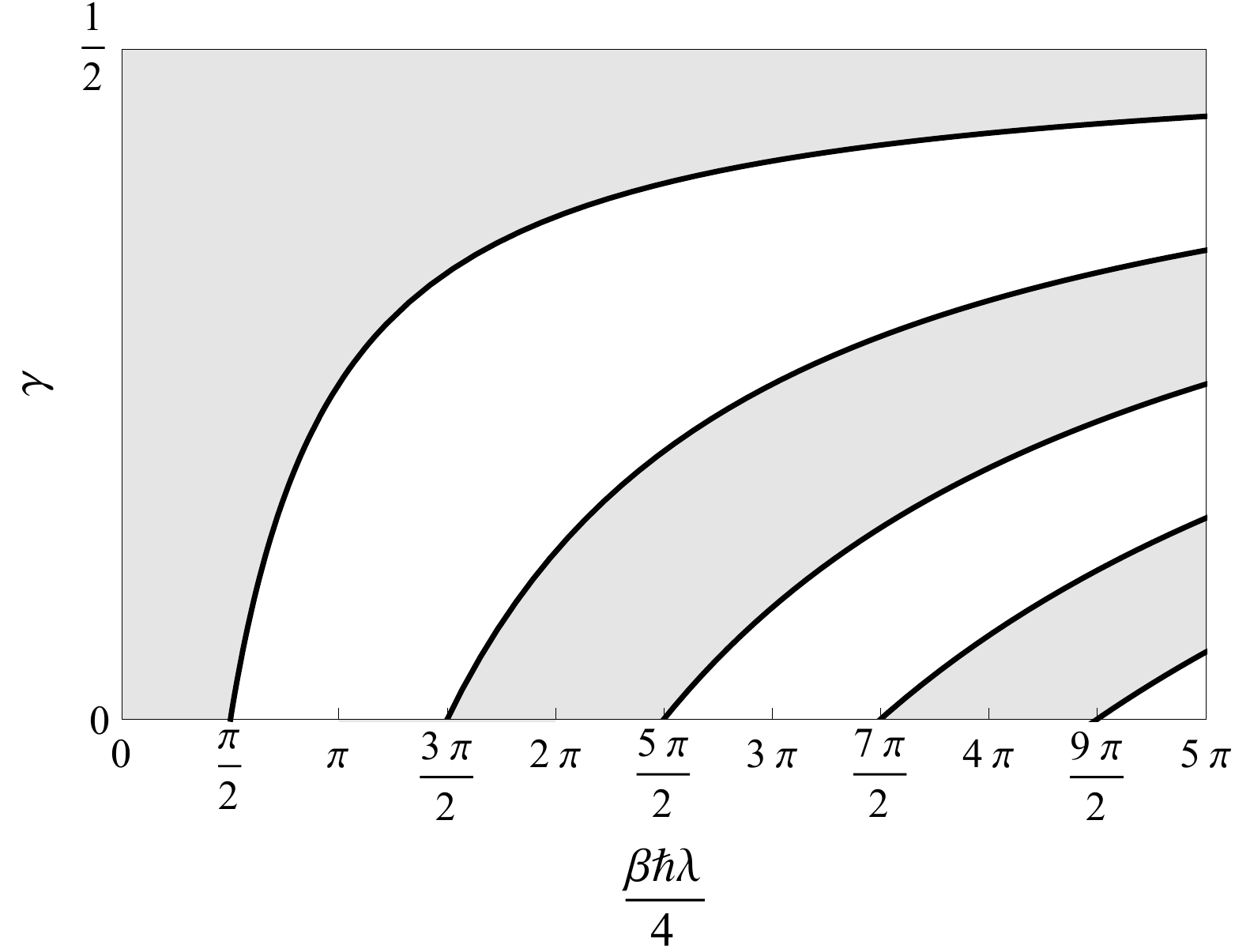}
\caption{Region (shaded by gray) of $c_1(\gamma)\ge 0$.}
\label{lambda range}
\end{figure}

It is straightforward to prove the statement (iii) from the condition $c_1(\gamma)\ge 0$,
which is equivalent to
\begin{align}
\cos\left((1-2\gamma)\frac{\beta\hbar\lambda}{4}\right)
&\ge
0
\label{cos>=0}
\end{align}
for $0\le\gamma\le 1$. 
Since the condition (\ref{cos>=0}) is symmetric around $\gamma=\frac{1}{2}$,
it is sufficient to restrict ourselves to $0\le\gamma\le\frac{1}{2}$.
The allowed region of $(\lambda,\gamma)$ is depicted in Fig.~\ref{lambda range}.
The condition (\ref{cos>=0}) means that $\cos\theta\ge 0$ for $-\frac{\beta\hbar\lambda}{4}\le\theta\le\frac{\beta\hbar\lambda}{4}$.
Therefore the interval $[-\frac{\beta\hbar\lambda}{4},\frac{\beta\hbar\lambda}{4}]$
must be included in the interval $[-\frac{\pi}{2},\frac{\pi}{2}]$, which is satisfied
if and only if $\frac{\beta\hbar\lambda}{4}\le\frac{\pi}{2}$. This proves the inequality (\ref{lambda bound}).
$\Box$

{\it Alternative proof. ---}
There is another way to show the statements (i) and (ii) of the theorem with the use of analytic continuation.
As we have seen in the above, $F(z)$ is analytic in the region of $\Omega$, and on the real axis 
($z=t\in \Omega\cap\mathbb R$) $F(z)$ is given by Eq.~(\ref{F_1/2(t)}).
If (a) there exists an asymptotic expansion of $F(z)$ with respect to $\epsilon$ up to $O(\epsilon^2)$, 
and if (b) each term in the expansion
is also analytic in the region of $\Omega$,
then due to the uniqueness of analytic continuation we obtain
\begin{align}
F(z)
&=
c_0-\epsilon c_1 e^{\lambda z}+O(\epsilon^2)
\label{F(z)}
\end{align}
for $\forall z\in \Omega$. Substituting Eq.~(\ref{F(z)}) in Eq.~(\ref{F_gamma(t) 2}) gives
\begin{align}
F_\gamma(t)
&=
c_0-\epsilon c_1 {\rm Re} \left[e^{\lambda(t+i(\gamma-\frac{1}{2})\frac{\beta\hbar}{2})}\right]+O(\epsilon^2)
\notag
\\
&=
c_0-\epsilon c_1 \cos\left((1-2\gamma)\frac{\beta\hbar\lambda}{4}\right)e^{\lambda t}+O(\epsilon^2),
\end{align}
which is equivalent to (i) and (ii).

The remaining task is to show that the assumptions (a) and (b) made in the above argument
are true. To this end, we first need to show that $\lim_{\epsilon\to 0} F(z)$ exists and
that $\lim_{\epsilon\to 0} F(z)$ is holomorphic (note that $F(z)$ itself is holomorphic in $\Omega$ for each fixed $\epsilon$). 
Let us recall that
the real part of $F(z)$ is $F_\gamma(t)$, which converges uniformly in the region $\Omega$ in the limit of $\epsilon\to 0$.
Since $F(z)=F_\gamma(t)$ for ${\rm Im}\, z=0$, $F(z)$ converges
in the limit of $\epsilon\to 0$ for $z\in \Omega\cap\mathbb R$.
Now we invoke the following mathematical fact in complex analysis \cite{Rudin}:
Suppose that $f_n(z)$ ($n=1,2,3,\dots$) is holomorphic in the region $\Omega$,
$u_n(z)$ is the real part of $f_n(z)$, $\{u_n(z)\}$ converges uniformly on any compact subset of $\Omega$,
and $\{f_n(z)\}$ converges for at least one $z\in\Omega$. Then $\{f_n(z)\}$ converges
uniformly on any compact subset of $\Omega$. From this fact, it follows that $F(z)$ converges uniformly
in the region $\Omega$ in the limit of $\epsilon\to 0$. Uniform convergence guarantees that
$\lim_{\epsilon\to 0}F(z)$ is holomorphic in $\Omega$. By analytic continuation, we obtain
$\lim_{\epsilon\to 0}F(z)=c_0$ for $z\in\Omega$. We repeat the same argument with $F(z)$
replaced by $(F(z)-c_0)/\epsilon$, showing that $(F(z)-c_0)/\epsilon$ converges uniformly in $\Omega$
in the limit of $\epsilon\to 0$ and that $\lim_{\epsilon\to 0} (F(z)-c_0)/\epsilon$
is holomorphic in $\Omega$. Thus the assumptions (a) and (b) are shown to be true,
and the proof of the theorem is completed. $\Box$

\section{Discussions}
\label{discussion}

The assumption of the form of $F_\gamma(t)$ in Eq.~(\ref{asymptotic expansion})
for all $\gamma$ in $0\le\gamma\le 1$ is too strong for the purpose of showing (i) and (ii). 
As we have seen above, $F_\gamma(t)$ is uniquely determined
from $F_{\frac{1}{2}}(t)$. 
In fact, to prove (i) and (ii), it is sufficient 
to adopt a weaker assumption that Eq.~(\ref{asymptotic expansion}) holds for $\gamma=\frac{1}{2}$
and that there exists a uniform asymptotic expansion of $F_\gamma(t)$ in $D$.
If one further assumes $c_1(\gamma)\ge 0$ for $0\le\gamma\le 1$, then (iii) follows.
Also, the assumption of the uniformity of the asymptotic expansion seems to be rather technical.
Instead of uniformity, it is sufficient to assume (a) and (b) from the beginning
in order to prove the statements (i), (ii), and (iii) of the theorem.

Let us emphasize that in proving the theorem we cannot use the mathematical result employed in Ref.~\cite{MaldacenaShenkerStanford2016}: If $f(t+i\tau)$ is analytic in the half strip
$\{(t,\tau)|t>0, -\frac{\beta}{4}\le\tau\le\frac{\beta}{4}\}$, $f(t)$ is real for $\tau=0$, and 
$|f(t+i\tau)|\le 1$ in the entire half strip, then it follows that
\begin{align}
\frac{1}{1-f} \left|\frac{df}{dt}\right|
\le
\frac{2\pi}{\beta\hbar}+O(e^{-4\pi t/\beta\hbar}).
\label{inequality f}
\end{align}
It is argued in Ref.~\cite{MaldacenaShenkerStanford2016} that 
the appropriately normalized OTOC $f$ satisfies the assumptions of the above statement
if one assumes a factorization of certain time-ordered functions. From the inequality (\ref{inequality f}),
one can see that the exponential growth rate of $f$ is bounded by $2\pi/\beta\hbar$.
Here we cannot use this mathematical result simply because
the theorem does not assume anything about the behavior of $F_\gamma(t)$
out of the region $D=\{(t,\gamma)|t_1\le t\le t_2, 0\le\gamma\le 1\}$.
Thus it is impossible to bound $|F_\gamma(t)|$ in an entire region of a certain half strip
such as $\{(t,\gamma)|t\ge t_1, 0\le\gamma\le 1\}$ in our case.

If $\lambda$ were to exceed the bound $2\pi/\beta\hbar$, something strange would happen.
From the theorem, one can see that there exists some $\gamma$ in $0\le\gamma\le 1$ such that $c_1(\gamma)<0$.
This means that there exists an OTOC $F_\gamma(t)$ in the one-parameter family
that grows exponentially in the direction {\it opposite} to the one in which the initial-perturbation sensitivity grows.
That is, the direction of the exponential growth depends on the choice of the regularization
of $\langle [\hat A(t),\hat B(0)]^2\rangle$.
Although such a case is not excluded by the theorem, the exponential growth of the OTOC becomes 
regularization dependent, and is no longer universal.
As long as the exponential growth is universal, the growth rate must be bounded by the theorem.

The theorem can be extended to cases in which there is a subleading correction to the exponential growth
in the $O(\epsilon)$ term in Eq.~(\ref{asymptotic expansion}):
$F_\gamma(t)=c_0(\gamma)-\epsilon [c_1(\gamma)e^{\lambda(\gamma) t}+f_\gamma(t)]+O(\epsilon^2)$.
Here $f_\gamma(t)$ represents a subleading correction such as $c_2(\gamma)e^{\lambda'(\gamma) t}$
with $\lambda'(\gamma)<\lambda(\gamma)$ for $0\le\gamma\le 1$. 
By applying the same argument as in the proof of the theorem,
one obtains $F_\gamma(t)=c_0-\epsilon [c_1\cos\big((1-2\gamma)\frac{\beta\hbar\lambda}{4}\big)e^{\lambda t}+f_\gamma(t)]+O(\epsilon^2)$. As long as one requires the positivity of
the coefficient of the leading exponentially growing term
(i.e., $c_1(\gamma)\ge 0$), the exponent $\lambda$ in the leading term is bounded as in (\ref{lambda bound}).
Adding a subleading correction to the $O(\epsilon^0)$ term is also possible with the results unchanged.

The theorem does not exclude the growth of the OTOC faster than the exponential
such as $e^{\lambda t^2}$. Originally,
it has been conjectured \cite{MaldacenaShenkerStanford2016} that
\begin{align}
\frac{d}{dt}(F_d-F(t))
&\le
\frac{2\pi}{\beta\hbar}(F_d-F(t)),
\end{align}
where $F_d$ is a constant which $F(t)$ approaches after the dissipation time.
This is stronger than our statement that assumes an exponential growth from the beginning.
However, our argument in the proof of the theorem can be used to strongly constrain rapid growth of the OTOC. 
For example, if $F(t)$ takes a form of
$F(t)=c_0-\epsilon c_1 e^{\lambda t^2}+O(\epsilon^2)$
for $t_1\le t\le t_2$,
then a similar argument shows that 
$F_\gamma(t)=c_0-\epsilon c_1(\gamma,t)e^{\lambda t^2}+O(\epsilon^2)$ 
with $c_1(\gamma,t)=c_1 e^{-\lambda(1-2\gamma)^2(\frac{\beta\hbar}{4})^2}
\cos\big((1-2\gamma)\frac{\beta\hbar\lambda}{2}t\big)$.
That is, $F_\gamma(t)$ not only grows as $e^{\lambda t^2}$ but also oscillates with $t$.
If the duration of the growth $t_2-t_1$ is sufficiently large (i.e., $t_2-t_1>\frac{2\pi}{\beta\hbar\lambda}$), 
the $O(\epsilon^1)$ term of some of $F_\gamma(t)$ in $0\le\gamma\le 1$ must change the sign.
Thus it is impossible that 
all the members of the OTOCs in the one-parameter family grows as $e^{\lambda t^2}$
to the ``correct'' direction (such that the initial perturbation-sensitivity grows) for a sufficiently long-time duration.
The extension of the argument to other cases
including $e^{\lambda t^n} (n\ge 3)$ is straightforward.

Finally, let us point out that the theorem can be generalized to higher even-order OTOCs.
We define higher-order generalization of the one-parameter family
of the OTOCs $F_\gamma(t)$ (\ref{one-parameter family}) as
\begin{align}
F_\gamma^n(t)
&\equiv
\frac{1}{2}
{\rm Tr}\left(
\left[\hat\rho^{\frac{1-\gamma}{2n}}\hat A(t)\hat\rho^{\frac{\gamma}{2n}}\hat B(0)\right]^{2n}
\right)
\notag
\\
&\quad
+\frac{1}{2}
{\rm Tr}\left(
\left[\hat\rho^{\frac{1-\gamma}{2n}}\hat B(0)\hat\rho^{\frac{\gamma}{2n}}\hat A(t)\right]^{2n}
\right)
\end{align}
with $0\le\gamma\le 1$ and $n=1,2,3,\cdots$.
We note that $F_\gamma^n(t)=F_\gamma(t)$ for $n=1$,
$F_\gamma^n(t)$ is real for arbitrary $n$ and $\gamma$,
and $F_\gamma^n(t)$ is the $(AB)^{2n}+(BA)^{2n}$ part of the regularized
$\langle [\hat A(t),\hat B(0)]^{2n}\rangle$.
Again $F_\gamma^n(t)$ has appeared in the left-hand side (``fluctuation'' part) of the 
$2n$th-order out-of-time-order fluctuation-dissipation theorem
\cite{TsujiShitaraUeda2016},
\begin{align}
&
\sum_{\alpha_1,\alpha_2,\dots,\alpha_{2n}=\pm}^{\alpha_1\alpha_2\cdots\alpha_{2n}=+}
C_{[A,B]_{\alpha_1}[A,B]_{\alpha_2}\cdots[A,B]_{\alpha_{2n}}}^\gamma(\omega)
\notag
\\
&=
\coth\left((1-2\gamma)\frac{\beta\hbar\omega}{4n}\right)
\sum_{\alpha_1,\alpha_2,\dots,\alpha_{2n}=\pm}^{\alpha_1\alpha_2\cdots\alpha_{2n}=-}
C_{[A,B]_{\alpha_1}[A,B]_{\alpha_2}\cdots[A,B]_{\alpha_{2n}}}^\gamma(\omega),
\label{2nth-order FDT}
\end{align}
where $C_{[A,B]_{\alpha_1}[A,B]_{\alpha_2}\cdots[A,B]_{\alpha_{2n}}}^\gamma(\omega)$
is the Fourier transform of
\begin{align}
&
C_{[A,B]_{\alpha_1}[A,B]_{\alpha_2}\cdots[A,B]_{\alpha_{2n}}}^\gamma(t,0)
\notag
\\
&\equiv
{\rm Tr}\left(
\prod_{i=1}^{2n}\left[
\hat\rho^{\frac{1-\gamma}{2n}}\hat A(t)\hat\rho^{\frac{\gamma}{2n}}\hat B(0)
+\alpha_i\hat\rho^{\frac{1-\gamma}{2n}}\hat B(0)\hat\rho^{\frac{\gamma}{2n}}\hat A(t)
\right]
\right)
\end{align}
with $\alpha_i=\pm$. $F_\gamma^n(t)$ is related to the left-hand side of Eq.~(\ref{2nth-order FDT}) via
\begin{align}
F_\gamma^n(t)
&=
\frac{1}{2^{2n}}
\sum_{\alpha_1,\alpha_2,\dots,\alpha_{2n}=\pm}^{\alpha_1\alpha_2\cdots\alpha_{2n}=+}
C_{[A,B]_{\alpha_1}[A,B]_{\alpha_2}\cdots[A,B]_{\alpha_{2n}}}^\gamma(t,0).
\end{align}

Since $(-1)^n\langle [\hat A(t),\hat B(0)]^{2n}\rangle$ is positive semidefinite, 
it is reasonable to expect that $F_\gamma^n(t)$ grows exponentially (if it does) to the positive (negative) direction
for even (odd) $n$. Thus we assume that $F_\gamma^n(t)$ has a uniform asymptotic expansion of
\begin{align}
F_\gamma^n(t)
&=
c_0^n(\gamma)+(-1)^n\epsilon c_1^n(\gamma)e^{\lambda_n(\gamma)t}+O(\epsilon^2)
\end{align}
in the region $D=\{(t,\gamma)|t_1\le t\le t_2\, (t_1\neq t_2), 0\le\gamma\le 1\}$
with $c_1^n(\gamma)\ge 0$ and $\lambda_n(\gamma)>0$ for $0\le\gamma\le 1$.
If $c_1^n(\gamma)$ is nonzero at least at one $\gamma$ in $0\le\gamma\le 1$, then,
due to the same argument as in Sec.~\ref{theorem}, we can prove that $\lambda_n(\gamma)$
does not depend on $\gamma$ (hence we write $\lambda_n(\gamma)=\lambda_n$) 
and the $\gamma$ dependence of $c_1^n(\gamma)$ is determined as
\begin{align}
c_1^n(\gamma)
&=
c_1^n\cos\left((1-2\gamma)\frac{\beta\hbar\lambda_n}{4n}\right)
\end{align}
with a positive constant $c_1^n$. In order for the coefficient $c_1^n(\gamma)$ to be positive semidefinite
for $0\le\gamma\le 1$,
$\lambda_n$ must satisfy the inequality
\begin{align}
\lambda_n
\le
\frac{2n\pi}{\beta\hbar}
=
\frac{2n\pi k_BT}{\hbar}.
\label{nth order bound}
\end{align}
This is a generalization of the MSS bound to the higher-order OTOCs $F_\gamma^n(t)$.
If the bound (\ref{nth order bound}) is saturated, the dominant exponential growth
of the regularized $\langle [\hat A(t),\hat B(0)]^{2n}\rangle$ is given by 
$\exp(\frac{2n\pi}{\beta\hbar}t)=[\exp(\frac{2\pi}{\beta\hbar}t)]^n$.
This is natural since the fastest exponential growth of 
the regularized $\langle [\hat A(t),\hat B(0)]^2\rangle$
is given by $\exp(\frac{2\pi}{\beta\hbar}t)$.

To summarize, we have proved the inequality (\ref{MSS bound}) for the growth rate of the OTOCs
under the assumption that
all the OTOCs in the one-parameter family ($F_\gamma(t)$ with $0\le\gamma\le 1$) show a transient exponential growth
in the uniform asymptotic expansion
by using only the analytic properties of the OTOCs. We do not exclude the possibility
that some of the OTOCs in the one-parameter family might violate the MSS bound. However, in this case the sign
of the exponentially growing part depends on the regularization parameter,
which makes the exponential growth of the OTOC non-universal.
Our argument places a strong constraint on the growth of the OTOC faster than the exponential.
The obtained results are independent of the choice of the operators $\hat A$ and $\hat B$
and any details of the system, 
and applicable to arbitrary quantum systems in thermal equilibrium, including quantum black holes and 
strongly interacting many-body systems.

NT is supported by JSPS KAKENHI Grant No. JP16K17729.
TS acknowledges support from Grant-in-Aid for JSPS Fellows (KAKENHI Grant No. JP16J06936) and the Advanced Leading Graduate Course for Photon Science (ALPS) of JSPS.
MU acknowledges support by KAKENHI Grant No. JP26287088 and KAKENHI Grant No. JP15H05855.

\bibliographystyle{apsrev}
\bibliography{ref}

\begin{thebibliography}{35}
\expandafter\ifx\csname natexlab\endcsname\relax\def\natexlab#1{#1}\fi
\expandafter\ifx\csname bibnamefont\endcsname\relax
  \def\bibnamefont#1{#1}\fi
\expandafter\ifx\csname bibfnamefont\endcsname\relax
  \def\bibfnamefont#1{#1}\fi
\expandafter\ifx\csname citenamefont\endcsname\relax
  \def\citenamefont#1{#1}\fi
\expandafter\ifx\csname url\endcsname\relax
  \def\url#1{\texttt{#1}}\fi
\expandafter\ifx\csname urlprefix\endcsname\relax\def\urlprefix{URL }\fi
\providecommand{\bibinfo}[2]{#2}
\providecommand{\eprint}[2][]{\url{#2}}

\bibitem[{Lar()}]{LarkinOvchinnikov1969}
\bibinfo{note}{A. I. Larkin and Y. N. Ovchinnikov, Sov. Phys. JETP {\bf 28},
  1200 (1969).}

\bibitem[{Kit()}]{Kitaev2015}
\bibinfo{note}{A. Kitaev, talks at KITP (2015):
  \url{http://online.kitp.ucsb.edu/online/entangled15/kitaev/}, \url{http:
  //online.kitp.ucsb.edu/online/entangled15/kitaev2/}}.

\bibitem[{Mal()}]{MaldacenaShenkerStanford2016}
\bibinfo{note}{J. Maldacena, S. H. Shenker, and D. Stanford, J. High Energy
  Phys. 08 (2016) 106.}

\bibitem[{Sek()}]{SekinoSusskind2008}
\bibinfo{note}{Y. Sekino and L. Susskind, J. High Energy Phys. 10 (2008) 065.}

\bibitem[{She({\natexlab{a}})}]{ShenkerStanford2014a}
\bibinfo{note}{S. H. Shenker and D. Stanford, J. High Energy Phys. 03 (2014)
  067.}

\bibitem[{She({\natexlab{b}})}]{ShenkerStanford2014b}
\bibinfo{note}{S. H. Shenker and D. Stanford, J. High Energy Phys. 12 (2014)
  046.}

\bibitem[{She({\natexlab{c}})}]{ShenkerStanford2015}
\bibinfo{note}{S. H. Shenker and D. Stanford, J. High Energy Phys. 05 (2015)
  132.}

\bibitem[{\citenamefont{Sachdev and Ye}(1993)}]{SachdevYe1993}
\bibinfo{author}{\bibfnamefont{S.}~\bibnamefont{Sachdev}} \bibnamefont{and}
  \bibinfo{author}{\bibfnamefont{J.}~\bibnamefont{Ye}}, \bibinfo{journal}{Phys.
  Rev. Lett.} \textbf{\bibinfo{volume}{70}}, \bibinfo{pages}{3339}
  (\bibinfo{year}{1993}).

\bibitem[{Pol()}]{PolchinskiRosenhaus2016}
\bibinfo{note}{J. Polchinski and V. Rosenhaus, J. High Energy Phys. 04 (2016)
  001.}

\bibitem[{\citenamefont{Maldacena and Stanford}(2016)}]{MaldacenaStanford2016}
\bibinfo{author}{\bibfnamefont{J.}~\bibnamefont{Maldacena}} \bibnamefont{and}
  \bibinfo{author}{\bibfnamefont{D.}~\bibnamefont{Stanford}},
  \bibinfo{journal}{Phys. Rev. D} \textbf{\bibinfo{volume}{94}},
  \bibinfo{pages}{106002} (\bibinfo{year}{2016}).

\bibitem[{\citenamefont{Roberts and Stanford}(2015)}]{RobertsStanford2015}
\bibinfo{author}{\bibfnamefont{D.~A.} \bibnamefont{Roberts}} \bibnamefont{and}
  \bibinfo{author}{\bibfnamefont{D.}~\bibnamefont{Stanford}},
  \bibinfo{journal}{Phys. Rev. Lett.} \textbf{\bibinfo{volume}{115}},
  \bibinfo{pages}{131603} (\bibinfo{year}{2015}).

\bibitem[{Sta()}]{Stanford2016}
\bibinfo{note}{D. Stanford, J. High Energy Phys. 10 (2016) 009.}

\bibitem[{\citenamefont{Roberts and Swingle}(2016)}]{RobertsSwingle2016}
\bibinfo{author}{\bibfnamefont{D.~A.} \bibnamefont{Roberts}} \bibnamefont{and}
  \bibinfo{author}{\bibfnamefont{B.}~\bibnamefont{Swingle}},
  \bibinfo{journal}{Phys. Rev. Lett.} \textbf{\bibinfo{volume}{117}},
  \bibinfo{pages}{091602} (\bibinfo{year}{2016}).

\bibitem[{\citenamefont{Hashimoto et~al.}(2016)\citenamefont{Hashimoto, Murata,
  and Yoshida}}]{Hashimoto2016}
\bibinfo{author}{\bibfnamefont{K.}~\bibnamefont{Hashimoto}},
  \bibinfo{author}{\bibfnamefont{K.}~\bibnamefont{Murata}}, \bibnamefont{and}
  \bibinfo{author}{\bibfnamefont{K.}~\bibnamefont{Yoshida}},
  \bibinfo{journal}{Phys. Rev. Lett.} \textbf{\bibinfo{volume}{117}},
  \bibinfo{pages}{231602} (\bibinfo{year}{2016}).

\bibitem[{Yao()}]{Yao2016}
\bibinfo{note}{N. Y. Yao, F. Grusdt, B. Swingle, M. D. Lukin, D. M.
  Stamper-Kurn, J. E. Moore, and E. Demler, arXiv:1607.01801.}

\bibitem[{\citenamefont{Rozenbaum et~al.}(2017)\citenamefont{Rozenbaum,
  Ganeshan, and Galitski}}]{Rozenbaum2016}
\bibinfo{author}{\bibfnamefont{E.~B.} \bibnamefont{Rozenbaum}},
  \bibinfo{author}{\bibfnamefont{S.}~\bibnamefont{Ganeshan}}, \bibnamefont{and}
  \bibinfo{author}{\bibfnamefont{V.}~\bibnamefont{Galitski}},
  \bibinfo{journal}{Phys. Rev. Lett.} \textbf{\bibinfo{volume}{118}},
  \bibinfo{pages}{086801} (\bibinfo{year}{2017}).

\bibitem[{\citenamefont{Banerjee and Altman}(2017)}]{Banerjee2017}
\bibinfo{author}{\bibfnamefont{S.}~\bibnamefont{Banerjee}} \bibnamefont{and}
  \bibinfo{author}{\bibfnamefont{E.}~\bibnamefont{Altman}},
  \bibinfo{journal}{Phys. Rev. B} \textbf{\bibinfo{volume}{95}},
  \bibinfo{pages}{134302} (\bibinfo{year}{2017}).

\bibitem[{Pat({\natexlab{a}})}]{PatelSachdev2016}
\bibinfo{note}{A. A. Patel and S. Sachdev, arXiv:1611.00003.}

\bibitem[{Kur()}]{Kurchan2016}
\bibinfo{note}{J. Kurchan, arXiv:1612.01278.}

\bibitem[{Boh()}]{Bohrdt2016}
\bibinfo{note}{A. Bohrdt, C. B. Mendl, M. Endres, and M. Knap,
  arXiv:1612.02434.}

\bibitem[{Cho()}]{ChowdhurySwingle2017}
\bibinfo{note}{D. Chowdhury and B. Swingle, arXiv:1703.02545.}

\bibitem[{Pat({\natexlab{b}})}]{Patel2017}
\bibinfo{note}{A. A. Patel, D. Chowdhury, S. Sachdev, and B. Swingle,
  arXiv:1703.07353.}

\bibitem[{Tsu()}]{TsujiShitaraUeda2016}
\bibinfo{note}{N. Tsuji, T. Shitara, and M. Ueda, arXiv:1612.08781.}

\bibitem[{\citenamefont{Wigner and Yanase}(1963)}]{WignerYanase1963}
\bibinfo{author}{\bibfnamefont{E.~P.} \bibnamefont{Wigner}} \bibnamefont{and}
  \bibinfo{author}{\bibfnamefont{M.~M.} \bibnamefont{Yanase}},
  \bibinfo{journal}{Proc. Natl. Acad. Sci. U.S.A.}
  \textbf{\bibinfo{volume}{49}}, \bibinfo{pages}{910} (\bibinfo{year}{1963}).

\bibitem[{\citenamefont{Lieb and Robinson}(1972)}]{LiebRobinson1972}
\bibinfo{author}{\bibfnamefont{E.~H.} \bibnamefont{Lieb}} \bibnamefont{and}
  \bibinfo{author}{\bibfnamefont{D.~W.} \bibnamefont{Robinson}},
  \bibinfo{journal}{Comm. Math. Phys.} \textbf{\bibinfo{volume}{28}},
  \bibinfo{pages}{251} (\bibinfo{year}{1972}).

\bibitem[{\citenamefont{Nachtergaele et~al.}(2006)\citenamefont{Nachtergaele,
  Ogata, and Sims}}]{Nachtergaele2006}
\bibinfo{author}{\bibfnamefont{B.}~\bibnamefont{Nachtergaele}},
  \bibinfo{author}{\bibfnamefont{Y.}~\bibnamefont{Ogata}}, \bibnamefont{and}
  \bibinfo{author}{\bibfnamefont{R.}~\bibnamefont{Sims}}, \bibinfo{journal}{J.
  Stat. Phys.} \textbf{\bibinfo{volume}{124}}, \bibinfo{pages}{1}
  (\bibinfo{year}{2006}).

\bibitem[{\citenamefont{Hastings and Koma}(2006)}]{Hastings2006}
\bibinfo{author}{\bibfnamefont{M.~B.} \bibnamefont{Hastings}} \bibnamefont{and}
  \bibinfo{author}{\bibfnamefont{T.}~\bibnamefont{Koma}},
  \bibinfo{journal}{Comm. Math. Phys.} \textbf{\bibinfo{volume}{265}},
  \bibinfo{pages}{781} (\bibinfo{year}{2006}).

\bibitem[{Has()}]{Hastings2010}
\bibinfo{note}{M. B. Hastings, arXiv:1008.5137.}

\bibitem[{Rob()}]{RobertsStanfordSusskind2015}
\bibinfo{note}{D. A. Roberts, D. Stanford, and L. Susskind, J. High Energy
  Phys. 03 (2015) 051.}

\bibitem[{\citenamefont{Huang et~al.}(2016)\citenamefont{Huang, Zhang, and
  Chen}}]{Huang2016}
\bibinfo{author}{\bibfnamefont{Y.}~\bibnamefont{Huang}},
  \bibinfo{author}{\bibfnamefont{Y.-L.} \bibnamefont{Zhang}}, \bibnamefont{and}
  \bibinfo{author}{\bibfnamefont{X.}~\bibnamefont{Chen}},
  \bibinfo{journal}{Ann. Phys.}  (\bibinfo{year}{2016}).

\bibitem[{\citenamefont{Aleiner et~al.}(2016)\citenamefont{Aleiner, Faoro, and
  Ioffe}}]{Aleiner2016}
\bibinfo{author}{\bibfnamefont{I.~L.} \bibnamefont{Aleiner}},
  \bibinfo{author}{\bibfnamefont{L.}~\bibnamefont{Faoro}}, \bibnamefont{and}
  \bibinfo{author}{\bibfnamefont{L.~B.} \bibnamefont{Ioffe}},
  \bibinfo{journal}{Annal. Phys.} \textbf{\bibinfo{volume}{375}},
  \bibinfo{pages}{378 } (\bibinfo{year}{2016}).

\bibitem[{\citenamefont{Tsuji et~al.}(2017)\citenamefont{Tsuji, Werner, and
  Ueda}}]{TsujiWernerUeda2016}
\bibinfo{author}{\bibfnamefont{N.}~\bibnamefont{Tsuji}},
  \bibinfo{author}{\bibfnamefont{P.}~\bibnamefont{Werner}}, \bibnamefont{and}
  \bibinfo{author}{\bibfnamefont{M.}~\bibnamefont{Ueda}},
  \bibinfo{journal}{Phys. Rev. A} \textbf{\bibinfo{volume}{95}},
  \bibinfo{pages}{011601(R)} (\bibinfo{year}{2017}).

\bibitem[{Hae()}]{Haehl2016a}
\bibinfo{note}{F. M. Haehl, R. Loganayagam, and M. Rangamani,
  arXiv:1610.01940.}

\bibitem[{ana()}]{analyticity}
\bibinfo{note}{One can show this for a finite $N$ quantum field theory
  (including a quantum mechanical system) with finite volume
  \cite{MaldacenaShenkerStanford2016}.}

\bibitem[{Rud()}]{Rudin}
\bibinfo{note}{W. Rudin, {\it Real and complex analysis}, 3rd ed. (McGraw-Hill,
  New York, 1987), Chap. 11, Exercise 8.}

\end{thebibliography}

\end{document}